\documentclass[a4paper]{easychair}

\usepackage{doc}
\usepackage{color}
\usepackage{ragged2e}

\usepackage[]{hyperref}
\hypersetup{
    colorlinks=true,
}

\usepackage{graphicx}

\usepackage[utf8]{inputenc}
\usepackage{subfigure}
\usepackage{multirow}
\usepackage{url}

\usepackage{rotating}
\usepackage{pdflscape}

\title{Managing IEC 61850 Message Exchange for SDN-Controlled Cognitive Communication Resource Allocation in the Smart Grid}

\author{
Yanny Moscovits\inst{1}\thanks{Moscovits, Y. is with UNIFACS IPQoS research group}
\and
Eliseu Torres\inst{1}\thanks{Torres, E. is with FIBRE UNIFACS IPQoS research group}
\and
Joberto S. B. Martins\inst{1}\thanks{Prof. Dr. Martins, J. is with UNIFACS IPQoS and NUPERC research groups}
}

\institute{
  Salvador University,
  Salvador, Brazil\\
  \email{yannymoscovits@gmail.com, eliseutorres.gx@gmail.com, joberto.martins@gmail.com}
 }

\authorrunning{Moscovits and Martins}

\titlerunning{8th International Workshop on ADVANCEs in ICT, January~2020}

\begin{document}

\maketitle

\begin{abstract}
The IEC 61850 standard is being largely used in the Smart Grid (SG) context mainly due to its ability to address communication, interoperability and migration issues. IEC 61850 currently aims at internal substation communication. Nevertheless, there is a demand to generalize its use for distributed SG systems like Home Energy Management  Systems  (HEMS) and Advanced Monitoring Infrastructure (AMI) Communication which potentially involves distributed substations or distributed SG components. IEC 61850-based systems require constrained timing requirements for communication and the common approach is to allocate static link bandwidth resources leading in some cases to over dimensioning. This paper presents the Substation  Cognitive  Communication  Resource  Management (IC\textsuperscript{2}RM), that aims the management of bandwidth allocation for IEC messages using a cognitive approach for its provisioning and the SDN/OpenFlow for  its deployment. By dynamically deploying bandwidth for IEC messages, IC\textsuperscript{2}RM optimizes links between SG substations and systems and potentially reduces the operational costs (OPEX).
\end{abstract}




%
%


\section{Introduction}

The IEC 61850 standard is being largely used in the Smart Grid (SG) context. It addresses and standardizes important aspects of the grid operation and management such as the communication messages required, topologies and services. The IEC 61850 also proposes complete models, describing everything from physical devices to data and attributes \cite{honeth_application_2011} \cite{lopes_smart_2012}.



In the Smart Grid a robust and largely distributed network communication capability is essential to guarantee the grid operation and management \cite{bezerra_framework_2013}. The grid structure and operation can be segmented in various ways. This fundamentally depends on the problem perspective being focused. The most basic and generic structured segmentation consists of 3 general grid components that must communicate: i) generation; ii) transmission; and iii) distribution \cite{lopes_smart_2012}.

In regards to network communication supporting the main SG components, there are grid elements that have a functionality and communicate as part of their operation and management processes. Considering this perspective, the main functional elements communicating within the SG are: i) the Home Energy Management Systems (HEMS); ii) the Substation Automation Systems (SAS); iii) the Grid Energy Management System (GEMS); and iv) the Advanced Monitoring Infrastructure (AMI) Communication \cite{dorsch_analysing_2014}.


The Home Energy Management Systems (HEMS) is part of the Smart Grid on the consumption side where home appliances (e.g., air conditioner, dishwasher, dryer, refrigerator, kitchen stove, and washing machine) data are collected using smart meters. This data will be used to optimize power source and distribution. HEMS allows the end user to track consumption and optimize it, reducing energy costs \cite{niyato_machine--machine_2011}.

The Substation Automation Systems (SAS) enables control through physical elements without the need for human interference, increasing reliability and reducing the duration of disturbances or failures. For this, communication protocols between the IEDs (Intelligent Electronic Device) are used. There are approximately 150 different communication protocols for data transmission and 20 different communication protocols in specific equipment used in utility companies. This makes it difficult to interconnect equipment from different manufacturers and gateways introduce delays in messaging can lead to improper operation. In this specific SG functional component, IEC 61850 plays a fundamental role and is relevant.


The Grid Energy Management System aims the overall management of the entities involved in the SG such as distributed energy sources, microgrids, energy storage, smart buildings, smart homes and electric vehicles, among others.

The Advanced Monitoring Infrastructure (AMI) is a bi-directional communication network integrated with sensors, intelligent meters and monitoring systems that enable the collection and distribution of information between meters and utilities \cite{bouhafs_links_2012}.

Although the IEC 61850 was initially developed aiming to support internal substation communications addressing its problems and issues (SAS), the standard is being applied and extended for other SG communication scenarios like HEMS, GEMS and AIM \cite{hussain_iec_2018} \cite{ustun_iec_2019}.


The main advantages of using IEC 61850 in substations are its lower installation cost and its capability to support new features and advanced services such as the ones required in the Smart Grid. IEC 61850 defines external visible aspects of the devices beyond  data  encoding  on  the  wire and consequently enables interoperability, eases programming and lower equipment migration cost \cite{lopes_smart_2012}.

A problem concerning the adoption of IEC 61850 for communication either inside a substation or between substations is the need to have dedicated high speed capacity to support the timing requirement of priority messages. Inside a substation, dedicated switch ports are often allocated for IEC 61850 exchange of priority messages. For exchange of priority messages between substations, a dedicated overdimensioned link capacity is often used and this results in a relevant operational cost for the majority of the deployments. That being said, this paper  proposes the cognitive allocation of communication resources, like  link bandwidth and port capacity, in such a way that they can be optimized and shared among IEC 61850 messages.


This paper presents the Substation Cognitive Communication Resource Management (IC\textsuperscript{2}RM). IC\textsuperscript{2}RM objective is to allow the cognitive control of communication resources allocated for groups of IEC 61850 messages inside substation's network or between substations and functional elements of the Smart Grid. IC\textsuperscript{2}RM addresses the issue of optimizing network resources deployed for communication among Smart Grid functional elements.

In the next part of this article, Section \ref{sec:Problem} discusses IEC 61850 message types, their scope and related requirements. Section \ref{sec:relatedwork} indicates the relevant work being done related to SG components communication with IEC 61850.  Section \ref{sec:modeling} describes the IC\textsuperscript{2}RM architecture and 61850 data modeling approach used to control and manage the IEC messages. Section \ref{sec:ProofOfconcept} presents the implementation and section \ref{sec:conclusao} presents the final considerations.


\section{IEC Messages Types, Scope and Requirements} \label{sec:Problem}

The Substation Cognitive Communication Resource Management (IC\textsuperscript{2}RM) objective is to allow the cognitive control of communication resources allocated for groups of IEC 61850 messages and, in this perspective, its necessary to identify the messages types and scope alongside with their format and requirements.

There are 4 basic IEC 61850 messages types: i) Sampled Values (SV) messages; ii) PTP/SNTP (Precision Time
Protocol/ Synchronous Network  Time  Protocol) messages; iii) GOOSE (Generic Object Oriented Substation Event) messages; and iv) MMS (Manufacturing Message Specification) messages.


SV, PTP/SNTP and GOOSE IEC 61850 messages are transported over the SG network using either UDP/IP (User Datagram Protocol/ Internet Protocol) or straight in the Ethernet frame payload. SV messages are intended to support efficient monitoring and control of substation and SG equipment. PTP/SNTP messages are intended to support time synchronization among IEDs (Intelligent Electronic Device) or IEC-based equipment. GOOSE messages are mainly intended to support hard real-time control applications. The Manufacturing Message Specification (MMS) is a client-server based communication protocol. The client is a network application or device that requests data and actions from a server. The server contains a Virtual Manufacturing Device (VMD) in which it allocates objects and contents \cite{ozansoy_time_2008}.


The Smart Grid (SG) requires communication resources between components at various levels and uses various network and communication technologies \cite{lopes_smart_2012}. A relevant question concerning IEC 61850 in the SG is: What is the IEC 61850 main utilization scope and what are its deployment communication issues?

\begin{figure}[ht]
\centering
\includegraphics[width=.9\linewidth]{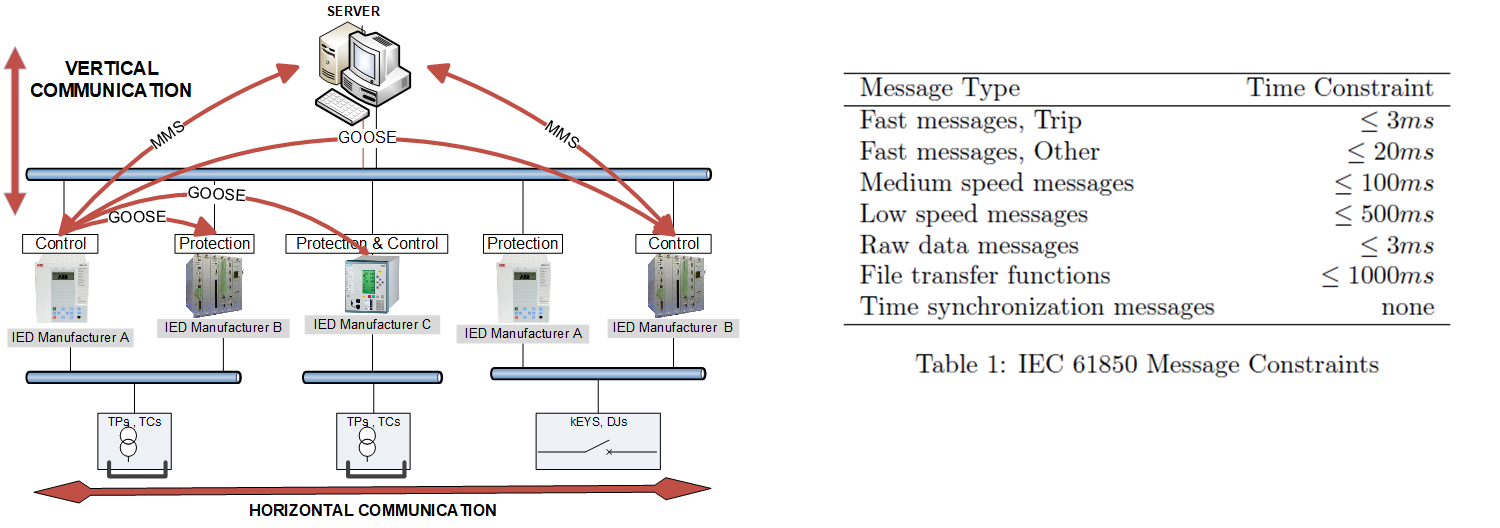}
\caption{IEC 61850 Horizontal and Vertical Communication Messages and Timings \cite{lopes_smart_2012}}
\label{fig:HVCommunication}
\end{figure}

The scope of the IEC 61850 in the Smart Grid has been primarily defined on supporting message exchanges internally and between substations. In transmission and distribution substations, the IEC 61850 supports two group of messages that provide communication with different functionality and timing requirements: i)  Horizontal communication; and ii) Vertical communication (Figure \ref{fig:HVCommunication}) \cite{lopes_smart_2012}


Horizontal communication uses typically GOOSE messages and are intended to support critical protection and control applications with real-time transmission delay requirements. Vertical communication uses typically MMS messages that are intended to support non-critical supervision applications. An example of typical timing requirements for the utilization of IEC 61850 message exchanges is illustrated in Table 1 (Figure \ref{fig:HVCommunication}) \cite{leon_real-time_2019}.



It is a fact that the main IEC 61850 communication approach to support the exchange of messages in substations is to over dimension network links in such a way that IEC messages always get all bandwidth they need and pass through with the required delay. The basic IC\textsuperscript{2}RM motivation is then to propose a new approach for link resource allocation looking for answers to the following research question: Is it possible to deploy IEC 61850 messages communication using links with shareable and limited resources?

The relevance of this approach is based on the following aspects: i) an IEC 61850 network with shareable resources do represent a more economical and efficient use of network technologies and resources in substations and among substations. In effect, the over dimensioning of substation's communication resources does represent an investment (CAPEX). On the other hand, the over dimensioning of communication resources among substations (typically wide area telecommunication links) does represent an operational cost (OPEX) and, as such, its reduction is relevant; and ii) SG uses multiple systems at substation level that require communication with heterogeneous requirements and the deployment of shareable  resources lead to a potentially more efficient solution.

The next rationale involving IEC 61850 message exchanges in the Smart Grid is: Can machine learning (ML) be used to support efficient and adequate communication resource allocation for IEC messages and other application and systems in substations and among substations?

\begin{figure}[ht]
\centering
\includegraphics[width=.5\linewidth]{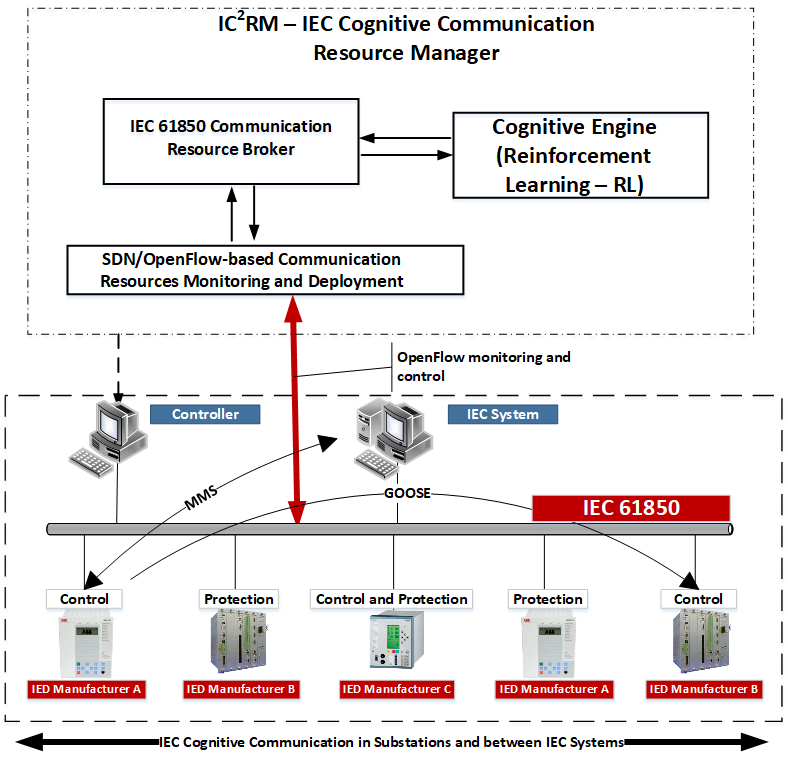}
\caption{IC\textsuperscript{2}RM Architecture and Operation Scope}
\label{fig:CognitiveCommunication}
\end{figure}

The IC\textsuperscript{2}RM is a cognitive communication approach based on SDN/OpenFlow for IEC 61850's communication resources allocation in the Smart Grid considering message exchanges inside substation and between substations and SG systems.  IC\textsuperscript{2}RM architecture is illustrated in Figure \ref{fig:CognitiveCommunication} and, in summary, it aims to allow the utilization of shareable communication resources by critical and non-critical IEC messages for functional components of the Smart Grid.

\section{Related Work \label{sec:relatedwork}}

Ustun (2019) in \cite{ustun_iec_2019}, presents recent application of the IEC 61850 to support communication in various Smart Grid application scenarios like substation communication, microgrid communication and home network communication, among others. Ustun (2011) in \cite{ustun_extending_2011} applies extensive communication capabilities of the IEC 61850 in microgrid distributed energy resource (DER) deployments with fault current limiters IEC modeling. Solar Home System (SHS) and Smart Meter(SM) are modeled in \cite{hussain_communication_2018} with IEC 61850 communication performance evaluation for different network technologies.

IEC 61850 communication-based coordinated operation of distributed energy resources (DER) and locally controlled distribution static compensators (DSTATCOM) is presented in \cite{hussain_iec_2018} where MMS type IEC 61850 messages are mapped onto the XMPP (eXtensible Message Presence Protocol) web protocol to provide microgrid communication.

In all the above references, no resource allocation scheme is presented for optimize IEC message communication and guarantee its time constraints. To the extent of our knowledge, IEC 61850 message time guarantees deployed using SDN and machine learning support for managing real time message constraints have not yet been proposed.

\section{IC\textsuperscript{2}RM Architecture and Link Management}

The IC\textsuperscript{2}RM architecture is illustrated in Figure \ref{fig:CognitiveCommunication}. IC\textsuperscript{2}RM is composed by 3 modules: i) The IEC 61850 communication broker; ii) a cognitive engine; and iii) a SDN-based network monitoring and communication resource deployment module.

The IEC 61850 communication broker manages the constrained and shareable communication resources either inside or between substations and systems. It decides how much bandwidth is allocated for each specific set of messages exchange.

The cognitive module processes all new message exchanges and, dynamically, suggests to the broker what resource allocation action should be executed upon the set of switches used in the substation or between substations. The main purpose of the cognitive engine is to provide feedback to the broker in terms of what is the best resource allocation action to be be executed based on the learning process involved in the system operation.

The SDN-based network monitoring and communication resources deployment module is basically an SDN interface between the broker and the set of OpenFlow-based switches. It monitors new incoming IEC 61850 message traffic by the use of OpenFlow \textit{Packet-In} messages and sets-up the flow-tables in the switches of the target communication grid system.

The IC\textsuperscript{2}RM's operation (dataflow) is modeled using the following principles: i) all IEC 61850 message exchanges are, at least during the first message exchange between control or supervisory equipment, inspected by the IC\textsuperscript{2}RM; ii) there is a message priority scheme defined by the manager that basically defines messages that do have stringent bandwidth reservation allocation and messages that do not need it; and iii) there is a message communication identification model in a way that messages can be identified and detected for on-the-fly processing.

\section{Message Identification Modeling for SDN-based Link Management supporting IEC 61850-based Systems} \label{sec:modeling}

IEC 61850-based systems are modeled using typically the following steps: i) An information model is proposed with the modules that communicate in the IEC 61850-based system; ii) A set of messages is created supporting the expected service or functionality for the system; and iii) These messages are mapped onto the basic set of IEC messages. As an example, Kikusato in \cite{ustun_iec_2019} defines an IEC-based information module with a set of messages for an EV (Electric Vehicle) charging system.

IC\textsuperscript{2}RM focuses on managing basic IEC 61850 messages and, as such, IC\textsuperscript{2}RM message identification modeling has to do with identifying and allocating link bandwidth for these messages. This approach guarantees that IC\textsuperscript{2}RM Message would work with any IEC 61850-based system development.


In terms of the IC\textsuperscript{2}RM implementation, the following priority message modeling is defined: i) GOOSE and SV messages have an assigned minimum private bandwidth allocated; ii) MMS and PTP/SNTP share a maximum limited amount of link bandwidth; and iii) MMS and PTP/SNTP maximum configured bandwidth can be, dynamically, re-allocated to GOOSE/ SV messages to guarantee its timing requirements. This priority modeling approach reflects the main objective of the IC\textsuperscript{2}RM which is to provide enough bandwidth to priority messages (GOOSE/ SV) while keeping some room to allow less priority messages (MMS and PTP/SNTP) over a restrained link with limited bandwidth resources.


From the operational point of view, IEC 61850 message resource allocation requires the following actions to be executed by the IC\textsuperscript{2}RM: i) GOOSE and SV messages exchanged by IEC-based equipment are processed by the broker when communication starts (1st message) to allocate bandwidth resource; ii) IC\textsuperscript{2}RM implements soft-state control of critical and non-critical message exchanges among all IEC-based equipment; and iii) Messages are identified using SDN/OpenFlow \textit{PacketIn} message and other OpenFlow protocol resources.

\begin{figure}[ht]
\centering
\includegraphics[width=.4\linewidth]{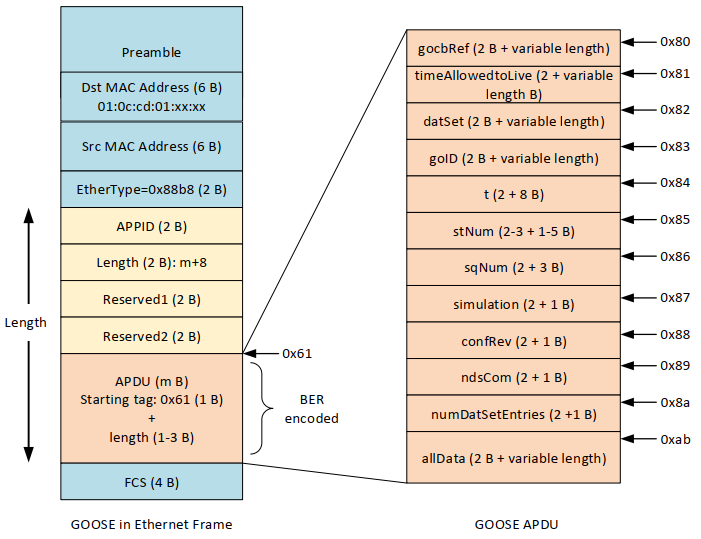}
\caption{IEC 61850 GOOSE Message Structure \cite{matousek_description_2019}}
\label{fig:GOOSEMessage}
\end{figure}

IC\textsuperscript{2}RM message identification modeling uses the following OpenFlow flowtable parameters:

\begin{itemize}
    \item GOOSE messages (Figure \ref{fig:GOOSEMessage}) \cite{matousek_description_2019}: i) Source/ Destination MAC addresses (equipment identification); ii) EtherType (GOOSE message); and iii) APPID (Application ID).
    \item SV messages: i) Source/ Destination MAC addresses (equipment identification); ii) EtherType (SV message); and iii) APPID (Application ID).
\end{itemize}



\section{IC\textsuperscript{2}RM Prototype Implementation} \label{sec:ProofOfconcept}

\begin{figure}[ht]
\centering
\includegraphics[width=.55\linewidth]{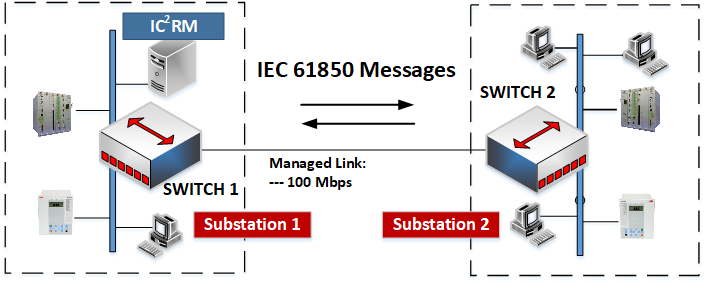}
\caption{Inter-Substation Communication Experimental Setup}
\label{fig:InterCom}
\end{figure}

The IC\textsuperscript{2}RM prototype implementation focuses initially on managing inter-substation IEC 61850 message exchanges as illustrated in Figure \ref{fig:InterCom}.

In this experimental setup 2 substations are interconnected using a  link (100 Mbps) and there is one ethernet OpenFlow-capable switch connecting IEC-capable and non-capable equipment per substation. The IC\textsuperscript{2}RM runs on a server (controller)  located in one of the substations. Its location in either substation 1 or 2 does not affect the overall operation because OpenFlow \textit{PacketIn} messages generated when new IEC 61850 message traffic is generated are forwarded to the controller independently of its position and both switches use the same controller.

This IEC 61850 application scenario is relevant to control and supervision applications that must operate beyond substations limits. This is, for instance, the case of microgird's communication in wide area for functionalities like the ones required by distributed energy resources (DER) which are highly dynamic in nature. Current IEC 61850 standard does not fully support this kind of setup and is oriented for internal substation communication.

The IC\textsuperscript{2}RM prototype uses the following software components: i) Ubuntu Server 14.04.4 as the operating system; iii) Mininet Version 2.2.2 to deploy the inter-substation communication setup of hosts and switches indicated in Figure \ref{fig:InterCom} iii) Oracle VM VirtualBox Version 4.1.18 r78361 as the virtualization software and iv) Pox controller; and v) OpenVSwich (OVS) as the SDN/Openflow basic components.

In terms of the IC\textsuperscript{2}RM's architecture module deployment, the IEC 61850 communication resource broker and the reinforcement learning (RL) cognitive engine are user applications running as a SDN-OpenFlow controller module in one of the hosts configured with the Mininet.

Link bandwidth allocation per prioritized (GOOSE and SV)  and non-prioritized IEC message exchanges is realized by using the Queue Manager \footnote {https://github.com/EliseuTorres/QueueManager} \cite{torres_bamsdn:_2018}. The QueueManager  is a module developed to allow the OpenFlow controller to manage dynamically the bandwidth allocation directly on OpenVSwitch. With this module, when the controller is setting a new flow, it can specify the maximum rate of bandwidth in a output port of OVS, and split the bandwidth between priority queues. Each queue receives then a specific tag that indicates the bandwidth rate. This solution solves a typical problem in controllers that can set queues, but the configurations of bandwidth allocation is intended to be done externally to the controller, preventing in some case the deployment of a dynamically configured switch.



\section{Final Considerations}
\label{sec:conclusao}

IC\textsuperscript{2}RM proposes a new scheme to allow priority and non-priority IEC 61850 messages to be exchanged in substations and between substations and SG functional components keeping their time constraint requirements. The IC\textsuperscript{2}RM uses a SDN-based broker that dynamically manages the bandwidth allocation for all messages flows. This allows link optimization and, consequently, the utilization of links that do not need  anymore to be over dimensioned, leading to the utilization of IEC 61850 standard in a larger set of distributed applications and systems in the Smart Grid context.

IC\textsuperscript{2}RM current prototype implementation runs on a Mininet emulated experimental setup with a POX SDN controller. The deployed prototype is able to detect new incoming IEC 61850 message flows that require bandwidth allocation using the OpenFlow \textit{PacketIn} resource. The broker allocates then the required bandwidth per message type to comply with its standard timing requirements. In current version of the IC\textsuperscript{2}RM prototype, a straightforward resource allocation method is used with the continuous and static allocation of bandwidth per message flow. At this prototype stage, the cognitive allocation approach based on the RL is not yet implemented since the current objective is to have a proof of concept of the prototype basic operation.

The next steps in the prototype implementation will include the use of reinforcement learning to learn about message traffic patterns that allow the dynamic management of under dimensioned links either inside or between substations.



\label{sect:bib}
\bibliographystyle{unsrt}
\bibliography{easychair}

\hfill \break

\vskip -2\baselineskip plus -1fil


\hfill \break


\end{document}